\documentclass[3p,times]{elsarticle}

\usepackage{ecrc}

\volume{00}

\firstpage{1}

\journalname{Nuclear Physics A}

\runauth{Author1 et al.}

\jid{nupha}

\jnltitlelogo{Nuclear Physics A}

\usepackage{graphicx}
\usepackage{amsmath,amssymb}

\begin{document}

\begin{frontmatter}



\title{Zeroing in on the initial state --- tomography using bulk, jets and photons}

\author[jyu,hip]{Thorsten Renk}
\author[vecc]{Rupa Chatterjee}
\author[jyu,hip]{Kari J. Eskola}
\author[jyu,hip]{Harri Niemi}
\author[jyu,hip]{Ilkka Helenius}
\address[jyu]{Department of Physics, University of Jyv\"{a}skyl\"{a}, P.O. Box 35, FI-40014 University of Jyv\"askyl\"a, Finland}
\address[hip]{Helsinki Institute of Physics, P.O. Box 64, FI-00014 University of Helsinki, Finland}
\address[vecc]{Variable Energy Cyclotron Centre, 1/AF, Bidhan Nagar, Kolkata-700064, India}




\begin{abstract}
One of the unsolved problems in the current 'standard model' of heavy ion physics is the apparent rapid thermalization of QCD matter in the pre-equilibrium stage. While it is challenging to probe this mechanism directly, there are now several observables available which allow tomographic imaging of the initial state geometry, which is expected to carry remnant information of the equilibration mechanism. On the fluid dynamics side, scaled fluctuations in the momentum space anisotropy parameters $v_n$ image the initial eccentricity fluctuations $\epsilon_n$ almost directly with only a weak dependence on the details of the fluid dynamical evolution. From a different direction, due to the strong non-linear dependence of their emission rates on temperature, thermal photons and their $v_n$ are very sensitive to the initial state graininess. Finally, the $v_2$ and $v_3$ of high $P_T$ hadrons coming from hard processes reflect the attenuation pattern of partons propagating through the inhomogeneous matter density after some fluid dynamical evolution. Combining information from all these channels does not yet lead to a fully consistent picture, however intriguing trends pointing towards non-trivial initial state dynamics emerge. 
\end{abstract}

\begin{keyword}
quark gluon plasma \sep initial state \sep tomography \sep jet quenching

\end{keyword}

\end{frontmatter}



\section{Introduction}
\label{intro}

One of the key concepts in discussing observables in the context of ultrarelativistic heavy-ion (A-A) collisions is tomography, i.e. the ability to image the density distribution of the QCD medium produced in such collisions as well as its time evolution. In particular, accurately imaging the density distribution at early times where a collective state of matter forms would allow to identify the physics processes underlying the equilibration process better.

Several different classes of observables have been demonstrated to have tomographic capabilities. The bulk matter itself evolves like a fluid, and consequently pressure gradients translate the initial spatial eccentricities (usually paramtetrized in terms of Fourier coefficients $\epsilon_n$) into final state momentum space anisotropies $v_n$, and it has been argued in \cite{EbyE} that if one scales the event-by-event (EbyE) fluctuations of $v_n$ by the event average, this is almost directly related to the EbyE fluctuations of $\epsilon_n$, thus allowing to image the medium at the very earliest times.

A measurement of thermal photon $v_n$ likewise is a tomographic probe of the early medium evolution. Since the electromagnetic coupling of photons to the medium is small, final state interaction is almost completely absent and thermal photon emission reflects the conditions at the time of photon emission. In particular, photon $v_n$ reflects the pressure-driven collective velocity (flow) field of the medium at the time of photon emission, which in practice predominantly probes the time window between equilibration and 3 fm/c \cite{Photon_v2}.

Finally, high $p_T$ parton production inside the QCD medium leads to a situation in which hard partons undergo a considerable energy degradation due to final state interaction as they propagate through the medium. Since the amount of modification of a parton shower in medium is on average proportional to the length and density of medium traversed, high $P_T$ hadron $v_n$ does not measure a pressure-driven flow but rather an extinction of yield in a given momentum window, i.e. directly images $\epsilon_n$ in position space. While it can be shown that the energy degradation of the leading parton peaks between 1-2 fm/c \cite{Edep}, it is also known that high $P_T$ $v_2$ has sensitivity to late time dynamics which may reach well into the hadronic phase around 5-7 fm/c \cite{SysHyd}.

Thus, there are various sets of tomographic probes available which all have somewhat different capabilities and probe different timeslices, and the aim of this work is to see whether combining information from these probes forms a coherent picture of initial state physics or not and whether these different probes constitute independent ways of measuring the same physics or are sensitive to rather different physics.

\section{Scaling relations}

In order to establish what properties of the medium evolution are constrained by current tomographic probes outside the context of a particular model, it is necessary to test a large parameter space of possible evolutions against the data. Until recently, this has been close to impossible due to the substantial numerical efforts needed (note however the recent large scale statistical analysis \cite{Fit}). However, the discovery of approximate scaling relations has opened a novel avenue to do such an analysis for a subset of observables. In \cite{EbyE}, it was realized that $\langle v_n \rangle \approx C_n \epsilon_n$ when averaging over many events with the same spatial eccentricity $\epsilon_n$ and that moreover $v_2 \approx C_2 \epsilon_2$ for individual events, i.e. the complications of the fluid dynamics expansion can to good accuracy be absorbed into a series of constants $C_n$ (note however that the relation is not precisely linear across the full centrality range \cite{Harri}). Thus, constructing observables in which the $C_n$ cancel, like $\delta v_2 = (v_2 - \langle v_2 \rangle)/ \langle v_2 \rangle$, allows to compare the EbyE final state probability of finding a $v_2$ in a centrality class  $P(v_2)$ directly to $P(\epsilon_2)$ in the initial state. Similarly, the evolution of $C_n$ with centrality is slow, i.e. for a reasonably narrow centrality range range from 0-5\%, $v_n = C_n \epsilon_n$ is also realized to good approximation. Finally, a similar argument also holds for the multiplicity production where $N_{final} = C N_{initial}$ parametrizes the amount of viscous entropy production during the evolution. Technically, utilizing such scalings corresponds to a dramatic decrease in processing time per event from $O(1)$ hour down to $O(0.1)$ seconds, allowing the desired rapid scan of parameter space.

Similar approximate scaling relations exist for some parameters influencing $v_2$ at high $P_T$. For instance, the response of $v_2$ a conjectured near $T_C$ enhancement of jet quenching over a baseline scenario \cite{Liao,NTC} is shown for different models of jet quenching and fluid dynamics in Tab.~\ref{T-1} and is found to be about 20\%, independent of the details of either parton-medium interaction. A similar statement can be made about the effect of viscous entropy production on $v_2$ over an ideal inviscid baseline, which yields about 50\% enhancement independent of the parton-medium interaction model used.   

\begin{table}[htb]
\begin{center}
\begin{tabular}{r|c}
model & NTC$/\epsilon^{3/4}$\\
\hline
ASW & 1.17\\
YDE 3d & 1.22\\
YDE 2d & 1.20\\
\end{tabular}
\hspace{1cm}
\begin{tabular}{r|c}
model & visc/ideal\\
\hline
ASW & 1.51\\
AdS & 1.44\\
YDE & 1.55\\
\end{tabular}
\end{center}
\caption{\label{T-1}Relative enhancement of high $P_T$ $v_2$ of charged hadrons in 30-40\% central 200 AGeV Au-Au collisions over a baseline, showing the effect of a near $T_C$ enhancement of jet quenching (left) and viscous corrections (right) for different the jet quenching models ASW \cite{ASW}, YaJEM-DE (YDE) \cite{YDE} and AdS \cite{AdS} in 2d and 3d viscous and inviscid fluid dynamics \cite{SysHyd}.}
\end{table}

Note however that not all relevant parameters factorize that way --- most prominently, the geometric size (which determines the probability density $P(L)$ of pathlengths $L$ partons travel through the medium) manifestly does not, as different parton-medium interaction scenarios have different dependences on $L$. As in the case of bulk matter, the scaling relations allow to quickly estimate the effect of e.g. viscous corrections without re-doing the full fluid dynamics stage.

\section{Bulk $v_n$ fluctuation tomography}

Utilizing scaling relations, we have performed a systematic analysis of $P(\delta v_2)$ for 0-20\% centrality range, $\langle v_2 \rangle$ and $\langle v_3 \rangle$ for the 0-5\% centrality range and multiplicity production across the full centrality range which is partially published in \cite{EbyE2}. Among the parameter variations we have explored are the degrees of freedom (DOF) assumed to be colliding (nucleons, constituent quarks, color charges), their initial distribution (Woods-Saxon, hard sphere, 2-dim disc), the entropy production mechanism (binary collisions (BC), wounded participants (WN)), the entropy production geometry (variable size of the entropy production region around the production point), non-linearities in entropy production (saturation models $\rho^\alpha$, $f \rho_{WM} + (1-f) \rho_{BC}$ with the local entropy density $\rho$, threshold saturation) and additional fluctuations (collision region size fluctuations, multiplicity production fluctuations per collision, random cell-by-cell fluctuations at different size scales).

The outcome of this investigation can be summarized as follows: Chiefly $P(\delta v_2)$ and $\langle v_2 \rangle$ probe the effective surface diffuseness of the produced matter density distribution, and to a lesser degree also the saturation of entropy production in the central overlap region, whereas the centrality dependence of multiplicity production only constrains saturation. Observables based on higher harmonics $P(v_3), \langle v_3 \rangle$ appear to fairly generic and do not constrain model assumptions specifically. Most surprisingly, other sources of fluctuations, e.g. in the entropy production per collision or superimposed sub-nucleon sized fluctuations are not constrained by any of the observables studied, only fluctuations in the geometry are constrained.

The findings indicate that the relevant DOF for entropy production are unlikely to be nucleons, as the data require a surface diffuseness of the medium which exceeds what a nucleon-based Glauber model can deliver, however sub-nuclon DOF (such as constituent quarks) naturally result in such increased blurriness. Saturation of entropy production appears a necessary ingredient to explain the data, however a saturation model such as $\rho^\alpha$ may be misleading, as it does not only affect the central high density region but also (unphysically) the diffuse surface which is probed by the observables.

\section{Jet tomography}

As discussed above, the observed $v_2$ at high $P_T$ is driven by the attenuation of hard probes due to final state interaction with the medium. Thus, any given value of $v_2$ is a measure of both the medium density evolution and the pathlength dependence of the attenuation physics \cite{SysHyd}. In order to turn this into a tomographic probe, this pathlength dependence needs to be constrained independently, which can for instance be done using jet-h or h-h correlations with a weaker sensitivity to assumptions about the medium \cite{Constraining}.  As established in \cite{SysHyd}, the geometrical size of the region in which jet and medium interact is the single most important factor determining the magnitude of $v_2$. Other factors (with their relative importance quoted) are the formation time of a thermalized medium (50\%), viscous entropy production (35\%) and differences in the density profile (15\%). Thus, jet tomography measures a rather different subset of medium properties than bulk $P(v_2)$ which is predominantly sensitive to the density profile. 

Adding a conjectured near $T_C$ enhancement of jet quenching \cite{Liao} into the parameter space, choosing the code YaJEM-DE \cite{YDE} for which the pathlength dependence is constrained by correlation observables and utilizing the approximate scaling laws discussed above, one finds allowed regions in the parameter space $(\tau_0, \eta/s, c)$ where $\tau_0$ is the equilibration time, $\eta/s$ the strength of viscous corrections in the hydrodynamical evolution and $c$ the parametrized strength of near $T_C$ enhancement \cite{NTC}. The $P_T$ dependence of charged hadron $v_2$ at high $P_T$ can be accounted for well by YaJEM-DE assuming either a large $\tau_0$ $> 1$ fm/c, or strong viscous entropy production $\eta/s > 0.2$, or substantial near $T_C$ enhancement, or various combination of these parameters with lower values. So far, no unambiguous determination of either parameter can be made

\section{Photon tomography}

Current computations of thermal photon $v_2$ in standard hydrodynamics lead to significant discrepancies with data and inclusion of pQCD photons further increases these discrepancies \cite{Photon_v2}. Possible reasons might reside with fluid dynamics being the wrong concept to describe the evolution (however note the framework is very successful for almost any other observable), with an underestimation of the initial spatial eccentricity of the emitting medium (however, as seen above, this is well constrained) or with photon emission rates (which however would have to be underestimated by more than an order of magnitude in the hadronic phase to account for the discrepancy to data).

An interesting observation possibly hinting at the solution has been made in \cite{Photon_v3} where it was realized that while a non-zero $\epsilon_3$ is necessary to generate photon $v_3$, the magnitude of $\epsilon_3$ is not a good predictor for the magnitude for photon $v_3$ which correlates much better with the strength of early time radial flow. This might suggest that the solution to the photon $v_2$ puzzle lies in pre-thermalization dynamics leading to some primordial flow (see e.g. \cite{Fries}).

If the pre-equilibrium physics has a lower photon emission rate than an equilibrated thermal system, then primordial flow would significantly increase photon $v_2$ and $v_3$. At the same time, late equilibration and primordial flow would also shift the peak of the parton-medium interaction strength to later times, thus also increase high $P_T$ $v_2$ and pushing the allowed range of $(\eta/s, c)$ into the more plausible region. Thus, so far photon tomography offers tantalizing circumstantial evidence for non-trivial initial state dynamics, but it is fair to say that the observable is currently not fully understood.

\section{Conclusions}

The analyses summarized here show that tomographic observables potentially probe quite a variety of medium properties, which in turn are related to interesting initial state and medium evolution physics. The surface diffuseness probed by $P(v_2)$ is very sensitive to the colliding degrees of freedom, whereas the core entropy production is related to saturation physics. Pre-equilibrium dynamics and the underlying classical field dynamics might be glimpsed through photon tomography, whereas jets have the capability to probe interesting late time dynamics, such as viscous entropy production or the change of medium properties near $T_C$.

So far, the different tomographic probes show at best hints of a consistent picture, which is both a curse and a blessing. On the positive side, this implies that the whole set of observables probes a large range of medium properties, however the disadvantage is that there is little capability to perform cross checks or more solid constraints by observing the same property independently in different analyses.

In general, currently the interplay between the various tomographic observables is subtle to understand and needs careful considerations across different observables to reduce systematic errors, but at the same time, they show an enormous potential for future progress in understanding many aspects of the dynamics of heavy-ion collisions.

\section*{Acknowledgements}

This work is supported by the Academy researcher program of the
Academy of Finland, Project No. 130472 and by Project No. 133005








\end{document}